# Social Dynamics: Parameter Estimate

Michael Zhdanov

Gainesville, Florida, U.S.A.

The goal of this paper is to estimate some parameters of simple social harmonic oscillations modelling U.S. Presidential Approval and Macropartisanship. The harmonic oscillations are simplest solutions of equations of the deterministic social dynamics (Zhdanov M., 2024), relating social bodies' positions in the Bayesian space of assessments to noncoercive driving forces acting on these social bodies. In some limits, such as individuals assessing each other, the space may be reduced to just sets of scores, ratings, or indexes used to measure social variables like rates of presidential approval or consumer sentiments. Thereby, relations are established between forces of social attraction and elasticity, considered in the presented paper, and assessments of social bodies including individuals, by other social bodies. The results are applied to estimate parameters of social bodies and their motions from survey data on Presidential Approval, U.S. Senator Approval, and Macropartisanship and provide one more perspective on some well-known issues like what moves macropartisanship and why presidential approval fluctuates.

“History repeats itself”, “Fashions always come back” are examples of well-known statements reflecting periodic processes frequently encountered in social life. Numerous illustrations in business include assets’ price and volume fluctuations. Assets may mean real estate, stocks, etc. They can be considered as “the work” element of the concept of transaction triangle (Mitchell, 2001, 2002) and offered by either “the creative entity” itself or by its agents at the respective real estate markets or stock exchanges.

The assets’ price and volume fluctuations depend on consumer sentiments (Asrian V., Fuchs W., and Green B., 2019). The same term, “consumer sentiments”, may also be used to describe non-business social phenomena such as macropartisanship and variability of presidential approval because the concept of transaction triangle is implicitly employed in some political and other social areas, despite that transaction prices in non-business fields are not as certain.

If there is a transaction triangle, there is also an eternal question, which can be rephrased as “To buy or not to buy”, and there are always two (not multiple) poles within any social body, even within a single individual, in accordance with the well-known Heigel philosophic principle of “Unity and Struggle of Opposites”. One of these “poles” is attractiveness of “the creative entity’s” offer or the “market power” (Danov M.A., Smith J.B., and Mitchell R.K., 2003). Another “pole” is always existing doubts causing the so-called social elastic force, acting in the opposite direction and trying to keep the existing state.

If individual voters or other voting social bodies (e.g., states in the case of the U.S. federal elections) must decide on “to buy or not to buy”, the forces affecting consumer/voter sentiments are elastic social forces. While market forces are short-term results of public speeches or other marketing events, the sentiments’ variability is due to free elastic oscillations, caused by these forces, with specific for every social body natural frequency, unless the market forces act continuously for extended time spans, making the respective motions to become forced ones.

The U.S. state governors' job performance ratings (Hansen S.B., 1999) are most likely free oscillations in the field of attraction with their own natural frequencies, fairly independent on the governors' efforts. This is a possible nature of what the author calls, "Life is Not Fair". If consumer social bodies don't have to decide on anything (e.g., within democracies in between elections or monarchies and autocracies any time), then consumer sentiments may still fluctuate in the field of attraction just because "people's moods change". The goal of this paper is to propose relations describing free social body oscillations in the fields of attraction and elasticity and to make approximate estimates of their parameters (see also Danov, 2024).

To achieve this goal, background concepts and laws of social dynamics (Zhdanov M., 2024) are briefly introduced here again in the first section called, "Deterministic Social Dynamics". Second-order ordinary differential equations, resulting from the above-mentioned concepts and laws, are presented in the following section, "Simple Social Harmonic Oscillations". Then some parameters of these solutions are estimated in the section called "Political Applications", where the presented theory is employed to analyze U.S. federal elections, macropartisanship, and variability of presidential approval.

## DETERMINISTIC SOCIAL DYNAMICS

Effects of driving forces on consumer sentiments, presidential approval, etc. become clearer through deterministic social dynamics (Zhdanov M., 2024). Its subjects, social bodies, are generalized consumers when they are parts of transaction triangles. A social body is defied as a set of interrelated individuals and/or other, smaller social bodies, consisting of individuals, who possess or manage together assets and/or share common values and goals. This definition means that “consumers” may also be organizational units and “corporations are people too” (Sepinwall A., 2019; Martin F., 2022). Simple social bodies are just individuals, while complex ones include more than one individual and, possibly, other social bodies.

To follow states of social bodies, their positions are defined in the Bayesian space of assessments, which is just a generalized way to represent results of measuring opinions and assessments through personal questioning and polls, expressed in various scores, ranks, ratings, indexes, etc. The space is composed of independent coordinates appropriate for any specific case.

The coordinates may be aimed to position professional or ethical assessments of social bodies, such as “the creative entities” in a transaction triangle, i.e. to display their skills, experiences, education, trustworthiness, attitudes towards family, democratic freedoms, nomocracy, and other common values. They may also be intended to assess products or services, “the work”, produced by the “creative entity”. Then the coordinates may be selected to assess their practical, utilitarian or aesthetic qualities or both. Therefore, depending on a specific situation, the space may be one, two, or multidimensional.

To acquire coordinates in this space, the evaluator social body should assess the subject in the range between -100 and 100 with the zero assigned to unknown subjects only. This range is chosen because such assessments could be interpreted as probabilities in the Bayesian sense that the said is true. For example, an assessment score of 90 of a presidential candidate’s generalized professional quality means the candidate would succeed professionally with the likelihood of 90%, and -90 means that the likelihood of failure is 90%.

The total social body's assessment, however, cannot be expressed in percentage units because an assessment by a complex evaluator social body is defined as sums of assessments by all social bodies composing the evaluator's complex social body, with positive and negative scores summarized separately. Then the range is virtually from - ∞ to + ∞, and the respective units are called "*leo*".

If both the evaluator and evaluated are simple social bodies, composed of just one single individual, such as a university professor and a student, then the Bayesian space is reduced to scores ranged from -100 to +100. If this is the case, a conventional universities' range between 0 and 100 may be considered a special case of the Bayesian space, where just half of the space range is used. Bayesian space is, in essence, a generalization of all other kinds of ratings, scores, indexes, if it can be reduced to them under appropriate constraints imposed on types of social bodies and assessment units.

A result of assessing one body by another one is a dot in the Bayesian space of assessments. If the evaluated body applies a force towards the evaluator, the same force of opposite direction is automatically applied to it, moving the dot in the Bayesian space. In the above-described case of a professor and a student, for example, this force is attractiveness of the student's answers to the professor. I.e., it is the "market power" (Danov M.A., et. al., 2003), generalized recently to become a "driving force" (Zhdanov M., 2024).

Amongst other forces, acting on social bodies, the most obvious one is the force of attraction. If this force did not exist, there would be no social animals themselves. It is a so-called "central" force (e.g., Kittel C., Knight W.D., and Ruderman M.A., 1973), which is inversely proportional to squared distanced between social bodies in the Bayesian space of assessments.

Another obvious force results from the Hegel principle of unity and struggle of opposites. Due to this principle, there should always be a force, opposing to any potential change in a body's state. The nature of this force is the power of doubts or contradictions, and it is called the force of social elasticity. This force is always directed inversely to any change, and it may often be considered linear in the first approximation. Therefore,

$$\boldsymbol{F}_e = - k_e \cdot \boldsymbol{r} \,. \qquad (1)$$

Here $\boldsymbol{F}_e$ is a vector of the social elasticity force, $\boldsymbol{r}$ is a vector of the body's displacement from a sustainable state, $k_e$ is yet an unknown coefficient of social elasticity in the Bayesian space of assessments, and the module $\boldsymbol{r}$ is a displacement $r$ from the sustainable state where the elasticity force $\boldsymbol{F}_e = 0$.

If there is no force applied to a social body, its coordinates in the Bayesian space will stay the same or continue moving with the same permanent speed and direction. This property is called inertia. The measure of inertia is mass. The greater a body's mass, the greater its inertia, and the greater should be a force capable of changing its state in the Bayesian space. For a complex social body, it should be dependent on how many social bodies compose the considered complex body and on their physical/military, intellectual/educational, and economic/assets/earnings powers.

Some of these powers don't matter in some specific circumstances. A state in the U.S., for example, is, obviously, a complex social body. Its military power is equal to zero because the defense function is delegated by states to the Federal Government. A very rough, very approximate estimate of a state's mass may disregard both intellectual/educational and economic/assets/earning powers as well as potentially large number of complex social bodies within the state, leaving just the simplest social bodies, the individuals. I.e., a state's mass may be considered equal to the size of the state's population or the number of its voters.

This is about the case of the U.S. federal, including presidential, elections, where the voters are, in essence, the U.S. states weighted by their population (the state's share of the total population of the United Sates). Therefore, a social body's mass is proportional to just its population in this simplified case.

Furthermore, mass is also a measure of social attraction, in addition to being a measure of inertia. Therefore, two bodies of masses $m_1$ and $m_2$ attract each other with a force, proportional to their product and inversely proportional, being a central force, to a squared distance between them in the Bayesian space of assessments:

$$F_a = \gamma \cdot m_1 \cdot m_2 / r^2. \qquad (2)$$

Here $F_a$ is a module of the force of attraction between two social bodies, $r^2$ is the squared distance between them in the Bayesian space of assessments, $m_1$ and $m_2$ are social masses of the two bodies, and $\gamma$ is a coefficient of attraction, independent on a particular body.

Two laws of social dynamics have been already introduced here as follows:

1) If there is no force applied to a body, it will stay at the same point in the Bayesian space of assessments or go on moving with a continuous speed and direction; and
2) If a body applies a force to another one, it will experience the same force itself, acting in the opposite direction.

Third and the major law relates a force, acting on a social body of mass $m$, and the body's motion in the Bayesian space as follows:

$$\boldsymbol{a} = \boldsymbol{F} / m\ . \qquad (3)$$

Here $\boldsymbol{F}$ is a vector of force, applied to a social body, $\boldsymbol{a} = d\boldsymbol{v}/dt$ is a vector of its acceleration, $\boldsymbol{v} = d\boldsymbol{r}/dt$ is a vector of its speed, $\boldsymbol{r}$ is a vector of the social body's position in the Bayesian space of assessments, and $t$ is time.

## SIMPLE SOCIAL HARMONIC OSCILLATIONS

Social oscillations caused by the force of elasticity (1) can be described in the first, linear approximation through combining it with the law of dynamics (3) to obtain a second order differential equation

$$m \cdot d^2 \boldsymbol{r} / dt^2 + k_e \cdot \boldsymbol{r} = 0, \tag{4}$$

which is reduced to a scalar equation

$$d^2 x / dt^2 + (k_e/m) \cdot x = 0 \tag{5}$$

for assessments depending on one dimensional variables only. Here $x$ is one of these variables, e.g., a professional or ethical dimension of a "creative entity" or "others" or a practical or aesthetic dimension of "work".

The equation (5) describes small social harmonic oscillations through its general solution

$$x = A \, Sin \, \omega t + B \, Cos \, \omega t \, , \tag{6}$$

where $\omega = \sqrt{k_e/m}$ is the circular frequency, $\omega = 2\pi / T$, with $T$ being the oscillations' period, $\nu = 1/T$ is the regular frequency, and, respectively, $\omega = 2\pi\nu$.

The coefficients $A$ and $B$ are determined from initial conditions at the moment of time $t = 0$. It is seen from (6) that

$$B = x(0), \tag{7}$$

and

$A = \mathrm{d}x/\mathrm{d}t(0){:}\omega = v(0)/\omega.$ (8)

Here $x(0)$ is a displacement of the body from the equilibrium at the moment of time $t=0$, and $v(0)$ is the speed of the body at the initial moment of time in the Bayesian space of assessments.

Despite described oscillations are the simplest, harmonic ones, they are very important because virtually any kind of observed periodic motion can be decomposed into a Fourier series, representing a sum of harmonic functions. They are free, autonomous motions, independent of a short-term force acting before they start and affecting just their amplitude through the initial conditions (7) and (8). Their frequency is their own, natural or resonant frequency. It doesn't depend on their amplitude in the linear approximation.

Another kind of considered motions is free oscillations in the field of the force of attraction (2). Because these forces are central, social bodies are shaped into balls in the Bayesian space of assessments. The balls are discs in a particular case of a two-dimensional space. Fig. 1 describes the attraction between a social body C and a large, complex body O in a two-dimensional Bayesian space of assessments with axis $x$ being ethical and $y$, professional dimensions. A disc centered at the point O with coordinates (0,0) is a large, complex social body, e.g., a large country, such as the U.S., consisting of a numerous number of smaller social bodies including the one centered at the point C.

Body C is located close to the radius $R$ of body O, and the forces of attraction between them are determined from the expression (2) with $r = R$, which is much greater than a module of assessment of an average, ordinary, publicly unknown individual or another small social body such as a small and unknown private company, NGO, etc. It means that body C is a public body relative to body O, i.e., it is known in person, by reference, or through media to many or all social bodies inside of body O.

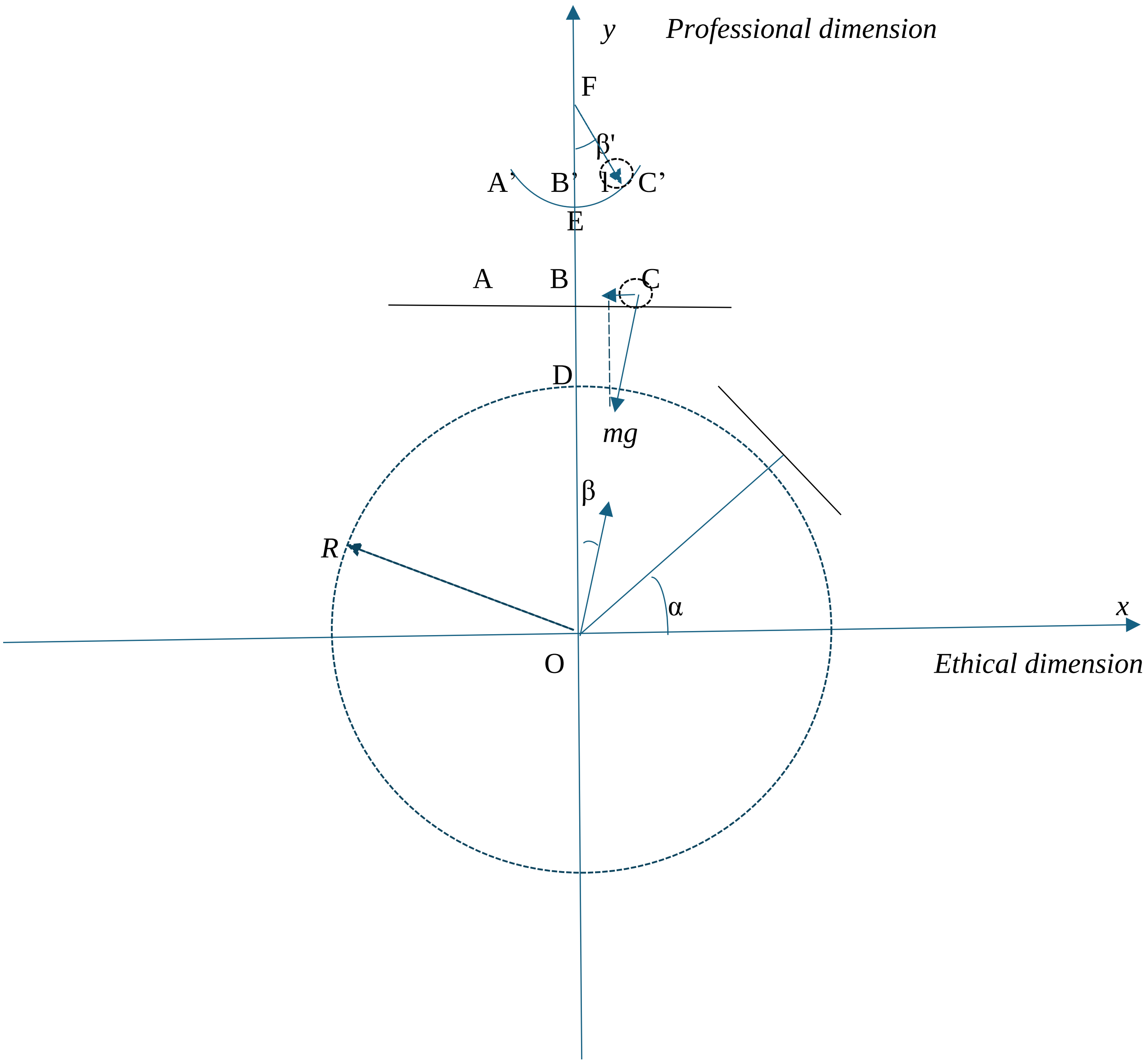

Figure 1. Social Attraction Oscillator

In considering the above-mentioned free oscillations, a particular example of body C is a widely known within the country public individual. Ideally, it is a former President of the U.S. or another inactive public social body, but an active body may be considered too if the individual is not a part of a transaction triangle, and its assessor don't have to decide on anything, but just want to perform an intermediate evaluation, e.g., between elections.

The area above the line $y$=OB is a locus of sets of the body's assessments relative to a large complex body O during a span of inactivity or regular intermediate work when the assessor doesn't have to make any decision, there is no transaction triangle, and the assessment reflects just sentiments of the evaluator changing with seemingly no reason.

Under an assumption that owing to the body C previous achievements, there is a straight line AC of a minimal professional assessment, which the inactive body can get, and the line is working similar, in a sense, to a rigid slippery plane laying on the Earth surface to limit physical motions of a material body. Despite that any social body within the body O can give any assessment, and some very limited number of these assessments may fall below this line, it is assumed, in essence, that eventual score obtained through summarizing evaluations by all social bodies will never fall below this line.

If this line is a curve, then resulting oscillations depend on its curvature. These oscillations are described below for a particular case of a circle A'EC' of radius $l$ depicted in Fig. 1 above line AC. The minimal professional assessment is given by bodies whose ethical assessment is zero, i.e. by those who don't know the subject from the ethical point of view in any way. With growing ethical evaluation, both positive and negative, the body's professional assessment score is increasing.

Depending on a range of evaluations of social bodies relatively to a much greater complex social body, the assessor, all smaller bodies are classified as being an "ordinary body", "ordinary public", and "outstanding public" social bodies. An "ordinary public" body is defined relatively to a large complex social body, such as a large country, as the one "laying on or near the surface" of the assessor. I.e., their modules of assessments are all within a certain layer BD, which thickness $\Delta R$ is much smaller than the radius $R$, i.e., $\Delta R \ll R$.

Therefore, ordinary public bodies are ones laying inside the segment [$R$, $R+\Delta R$], while "ordinary bodies" are all within (0, $R$), and "outstanding public" bodies are in the interval ($R+\Delta R$, ∞). Only small motions of the ordinary public bodies C and C' are considered, for which $Sin\beta$ = BC/CO << 1 for the straight line and $Sin\beta$'=B'C'/C'F<< 1 for the circle. I.e., the maximum displacement BC is much smaller than the radius $R$ and the maximum displacement B'C' is much smaller than the radius $l$.

If the radius $R$ of a complex body is defined, for example, as just the total number of social bodies withing the considered complex body, then an ordinary public body known to all these bodies will be assigned the total evaluation of $R$, laying right on the surface of the considered complex body, if modules of all of its assessments by every one of the compounding bodies are minimal, i.e., equal to one *leo*. If the complex body is, for example, a residential community, a radius, defined this way, is just the number of the community residents, their families, and associated workers, and it is also expressed in *leo*.

Due to the above definition of the ordinary public body, the values of assessments relative to social body O are about the same for all ordinary, not outstanding, public bodies, and they are close to $R$ for the example in Fig.1. This means that the mentioned ordinary public bodies are all in the layer BD << $R$. Therefore, distances are about the same between centers of the "ordinary" public bodies located within BD and the center of a large social body they all interact with. Then it is easy to obtain from (2) that the force acting on an ordinary public body of mass $m$ is equal to

$$P = g \cdot m \,, \qquad (9)$$

where $P$ is a force of social attraction acting on the body and directed towards the center of a the disc-like large complex social body O, such as a big and well-known country; $g = \gamma \cdot m_c / R^2$, with $\gamma$ being a constant coefficient in the expression (2), $m_C$ is a mass of the large social body O, $R$ is its radius, and $g$ is a "free-fall" acceleration, which is the same for all ordinary public bodies, whose assessments are all within the layer BD in the Fig. 1. This is a well-known fact for material bodies on Earth, and it is also true for ordinary public bodies interacting with a large social body of radius $R$, as far as $R = const$ with high accuracy due to the inequality BD << $R$ .

It is seen from Fig. 1 that a projection of the force of attraction $mg$ to the axis $x$ is equal to $-m \cdot g \cdot Sin\, \beta = -m \cdot g \cdot (x/R)$. The force is negative because its direction is opposite to the positive direction of the axes $x$. Then, due to the equation (3), $F = ma = m\, d^2x / dt^2 = -\, mg(x/R)$ and

$$d^2x / dt^2 + (g/R)\, x = 0 \, . \qquad (10)$$

This equation is the same as (5) except for the coefficient before x, and its solution is also the same expression (6) except that $\omega = \sqrt{g/R}$. It is seen that the value of $\omega$ is the same for all ordinary public bodies and depends on evaluator's parameters $g$ and $R$ only. This type of attraction oscillations is called here "universal". An equation, describing oscillations of body C' on the circle A'EC' is very easy to obtain too (e.g., Kittel C., et.al., 1973), and it is the following:

$$d^2x / dt^2 + (g/l)\, x = 0 \, . \qquad (11)$$

Its solution is the same relation (6), and its natural frequency is $\omega = \sqrt{g/l}$. This type of attraction oscillations is called here "pendulum-like" social oscillations.

## POLITICAL APPLICATIONS

### *U.S. Presidential Approval: Elastic Fluctuations*

Recent 2024 U.S. presidential elections revealed a significant discrepancy between results of public opinion polls, showing a very tight Harris-Trump race, and results of the elections with a landslide Trump's victory. The difference between the candidates exceeded well the declared 2% accuracy of these public opinion measurements.

In terms of the introduced here social dynamics, voters in the U.S. federal elections are social bodies, composing the federation, i.e., the U.S. states. Because the states differ significantly, a state's vote for a candidate should be weighed by the state's mass to ensure a fair contribution of the state towards the election results. A body's mass in social dynamics depends on three variables: intellectual, physical/military, and economic.

The state's military power is disregarded. It is virtually zero because they delegate their defense functions to federal authorities. Their economies differ significantly, and it might be fair to assign greater weights to states contributing more to the federation's prosperity. However, this is not the case now, and a state's economy is also disregarded in evaluating its mass or weight of its vote compared to other states.

Finally, a social body's mass should depend on its intellectual abilities, i.e. on the sum of all the state's social bodies such as individuals, universities, research laboratories, weighted by their education, ranks of its universities and its other private and public intellectual entities. In a limit of a small number of complex bodies compared to the number of simple ones - individuals - and small their total weighted contribution to the state's mass, obtain that the mass depends only on the number of individuals weighted by their education and/or other kinds of intellectual indicators. In one more limit, when all intellectual indicators may be considered about the same for all individuals or just intentionally disregarded, obtain that a state's mass is about equal to its population. This is the case nowadays: a state's vote is multiplied by its weight or mass, which is proportional to its population.

It has just been shown that weights used to differ the states' contributions to the U.S. federal elections now represent a limiting case (under some very rough assumptions) of the concept of mass in social dynamics. Only simplest social bodies, individuals, vote within the states, and "the winner gets it all". I. e., if 50% of a state's constituents plus one individual vote for a candidate, then another half of the state's voters is just disregarded. This may well be a major reason for the initially mentioned discrepancy between observed results of public opinion polls and results of the presidential elections.

In evaluating rates of presidential approval, a poll participant is asked, normally, to select one of three answers: approve, disapprove, or indifferent. The poll results are often presented as figures like Fig. 2 (Bidah S., et. al., 2020) depicting dependance on time of a percentage of the U.S. voters who approve or disapprove a president's work (Mr. Trump's one in this case).

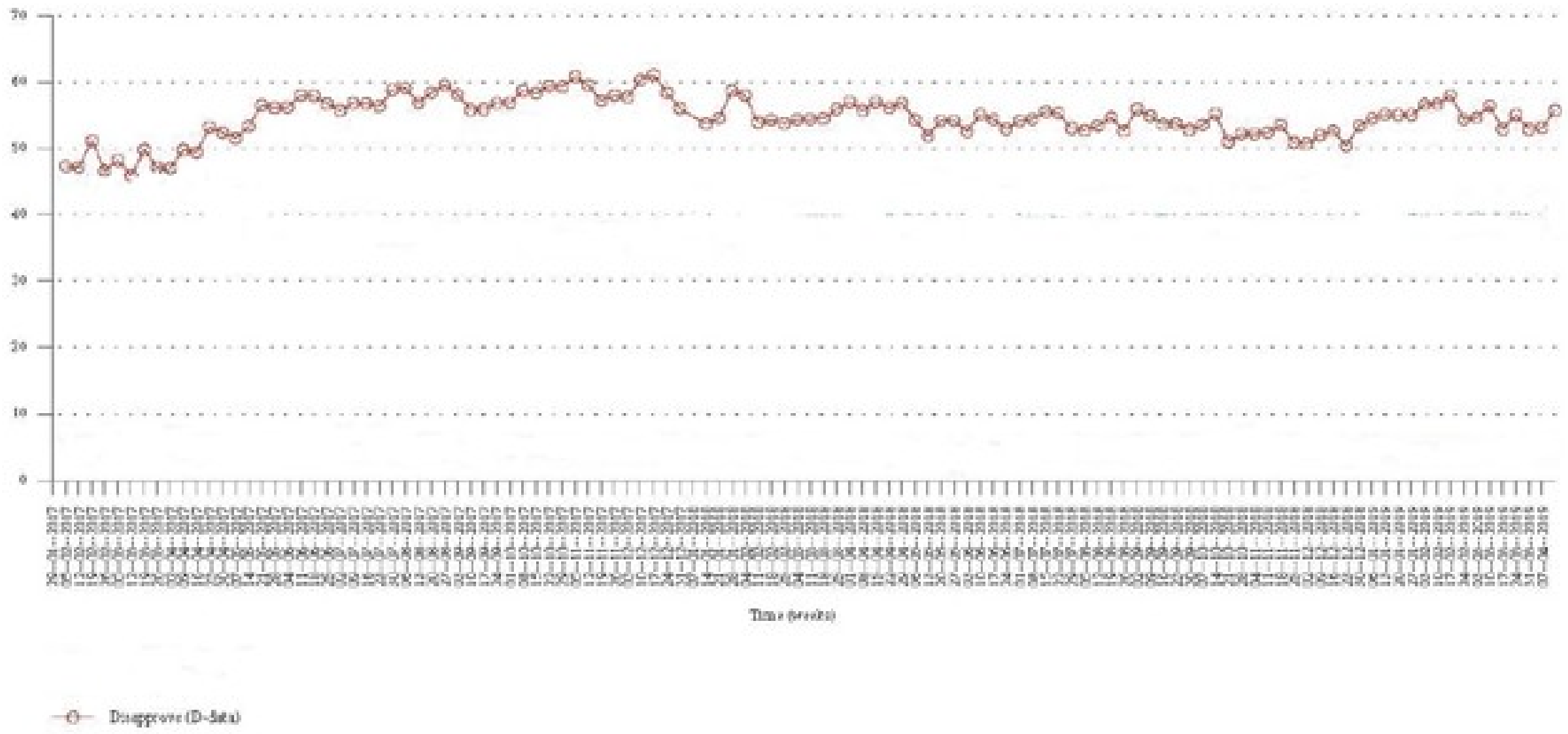

Fig. 2. Assessment of Donald Trump's professional performance.

To present the above results in the Bayesian space, all complex social bodies within the states should be disregarded and just the simplest ones, the individuals, should be considered with the numbers +1, -1, and 0 assigned to individuals, who responded “approve”, “disapprove”, or “indifferent” respectively.

Then both positive and negative numbers are to be summarized separately and made related to the total number of participants interviewed. Dependance on time of one of these numbers, extended to the total amount of the U.S. voters, is depicted in Fig. 2. This means the conventional approach of measuring public opinions through ratings is a particular case of the introduced here method, using the Bayesian space of assessments. Therefore, mathematical relations and concepts proposed in the first two sections of the present paper can all be used to interpret the variability, depicted in Fig. 2.

It is seen that the curve in Fig. 2 is packed with oscillations. Two periods of about 14 days between 19.02.17 and 19.03.17 are most likely elastic ones, unless they are due to an inaccuracy (e.g., just a weekly discreetness) of the calculations. The above two periods happened at the end of January 2017 just after Donald Trump started his first term. While the U.S. consumers didn’t have to decide on anything that time because they had already made their decision a few months before, the situation seems to be like the one when a person bought something in a store, came home, and is still having a last look at it again to see if it is a good deal. That was particularly important for Mr. Trump because he was not very experienced in politics at that time.

If these fluctuations are really elastic, then they obey to the equation (5) with $\omega_T=\sqrt{k_e/m_T}$, where $\omega_T$ is the circular frequency of the President Trump’s oscillations in the Bayesian space relative to the U.S. voters, $m_T$ is his social mass, and $k_e$ is the coefficient of elasticity of U.S. voters, which is the same for all presidents, serving at about the same time as well as for all other ordinary public bodies. Then

$m_T = k_e\, T^2/4\pi^2.$ (12)

A similar kind of ratings, obtained in the same way for another U.S. President Barak Obama, who served two terms in a row before President Trump's first term, is shown in Fig. 3 (Bidah S., et. al., 2020).

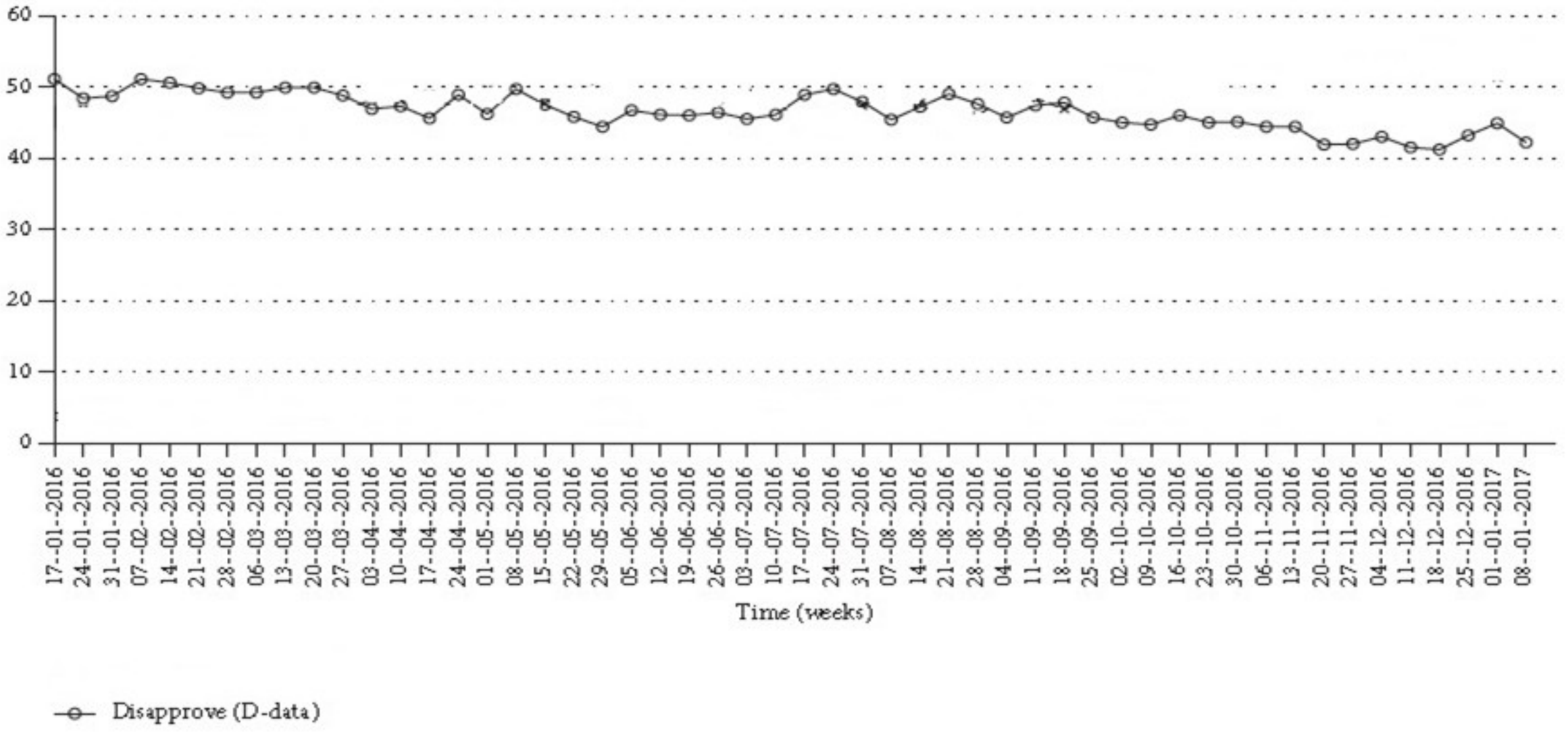

Fig. 3. Assessment of Barak Obama's professional performance.

There are several visible oscillations in this figure too with just two shortest periods of 14 days (from 10.04.16 to 08.05.16) compared to Trump's Fig. 2, where there are at least four of them in total. This fact becomes clear if the 14-day oscillations are elastic because consumer doubts in Obama should be lower: he is a professional politician who was serving his second term, and elastic oscillations are less frequently encountered compared to the case of Mr. Trump who was a businessman, successful entrepreneur, serving his first presidential term. The timing of the Obama's oscillations speaks also in favor of the "elastic version": April 2016 was about the time when Donald Trump's election campaign was already going on, and he is known to be a very persuasive, outstanding speaker.

If this is the case that both Trump's and Obama's 14-day oscillations are real and elastic, then Obama's and Trump's social masses are the same due to the relation (12) because their periods are equal. It is possible if, for example, their intellectual abilities are of the same order of magnitude and Obama's greater physical/health power is balanced out by Trump's greater personal economic power. Then the coefficient of elasticity, obtained from (12), is equal to $k_e = 4\pi^2 m_T / T^2 = 0.2\ m_T/\text{day}^2$. If these equal masses of Obama and Trump are considered a reference mass for ordinary public social bodies, which is equal to unity, then $k_e = 0.2/\text{day}^2$.

If the period of 14 days is not natural and is due to some calculational inaccuracies, including the observations' weekly discreetness, then next two periods to assume being elastic fluctuations are 21 days for Trump (13.05.18-24.06.18 in Fig. 2) and 28 days for Obama (10.07.16-02.10.16 in Fig. 3). Then it is easy to obtain from the relation (12) for Trump and the same relation for Obama, that $m_O = (T_O/T)^2 \cdot m_T = (28/21)^2\ m_T = 1.78 m_T$, where $m_O$ and $T_O$ are Obama's social mass and the period of his elastic oscillations respectively. That means it is better to be young and healthy than senior and wealthy. However, social obesity may also be a problem in politics.

Therefore, results of this subsection show that elastic oscillation parameters of ordinary public bodies, including U.S. presidents or presidential candidates, depend on the bodies' social masses, a coefficient of elasticity of the assessor and, to some extent, on "external" forces initiated the oscillations like shocking speeches and other provoking actions and events.

*Macropartisanship and Presidential Approval: Attraction Oscillations*

A concept of macropartisanship was introduced by MacKuen M.B., Erikson A.S., and Stimson J.A. in 1989. It may be treated as an "aggregate distribution of party identification" (Green D., Palmquist B., and Schickler E., 1998). In terms of the presented approach, complex social bodies composed of simple partisan bodies "share self conceptions and, therefore, are more stable in changing political circumstances" than both social groups composed of independent members, unrelated to any political party, and complex social bodies, such as the country itself, including all parts of the societal spectrum.

This is well seen in Fig. 4 (Green D., et, al., 1998), where the curves for Presidential Approval and Consumer Sentiments are significantly more ragged than the curve for Macropartisanship, fluctuating with much lower frequencies.

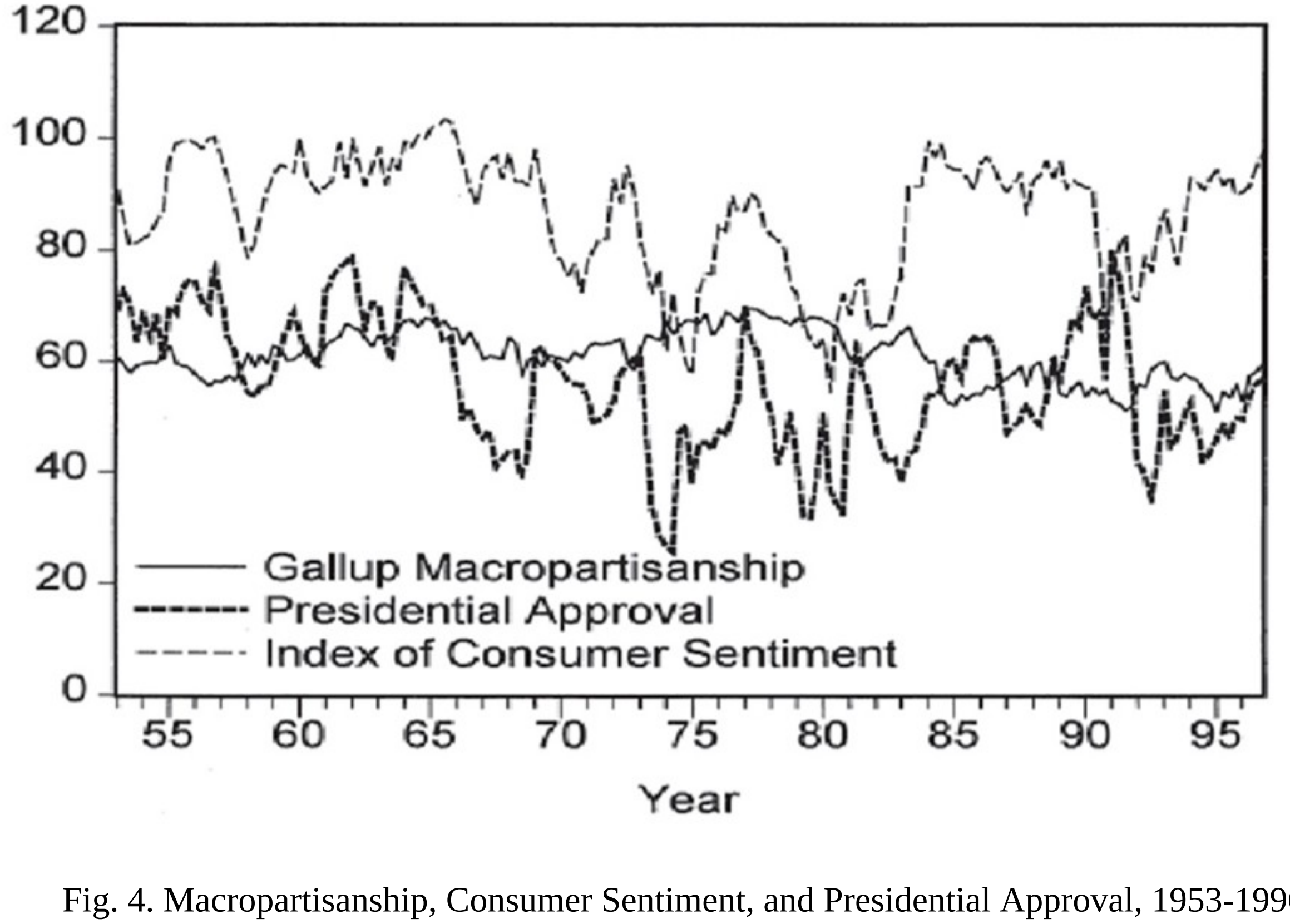

Fig. 4. Macropartisanship, Consumer Sentiment, and Presidential Approval, 1953-1996.

Although the amplitude of Macropartisanship oscillations is smaller, it is still possible to see three full-period low-frequency oscillations with minimums at about 1956, 1969, 1981, and 1994. Their period is about 13 years. Equations, derived here in the first two sections, can describe the variability of Presidential Approval and Consumer Sentiment because they refer to variability of assessments, specifically to the variability of assessments of ordinary public social bodies by a complex large social body such as a big country. Because almost a half of the U.S. voters are affiliated with political parties, macropartisanship itself is a kind of “people motions” (Tolstoy L., 2010, originally 1867), and the question, “What moves macropartisanship” is a part of the question “What is the nature of people motions”.

If the answer is the following: it is explicit or implicit assessments resulting in respective people’s actions, then Macropartisanship can also be described by social dynamics. These assessments include cumulative evaluations of a party leaders such as fluctuations of Presidential Approval, of voters’ sentiments regarding current and/or future economic situations under the ongoing and proposed leadership, i.e., the Consumer Sentiments or Expectations, and assessments of the so-called “deep states” on both sides, i.e., the most influential economically part of the evaluated party’s social bodies, which may not be even public themselves.

In spite of such a large number of Americans affiliated with political parties, people, while assessing any of the two major parties, evaluate not themselves, but a small number of the most influential party members, decision makers, and these evaluations of the Republican or Democratic party, which are both ordinary public social bodies, do satisfy the equations derived in the first two sections of the present paper. In practice, it might be more convenient to assess, for example, Republican or Democratic Senator Approval in stead of Republican or Democratic Party Approval if senators are tied in voters' minds to parties of their affiliations.

Fig. 5 (Anderson J.L. and Newmark A.J., 2002) shows Democratic and Republican Senator Approval from 1981 to 2000. The low-frequency minimums can be seen here too at about the same years of 1981 and 1994 as the above Macropartisanship's minimums. Therefore, it may be reasonable to assume that the U.S. partisanship approval varies with a period of about 10-13 years.

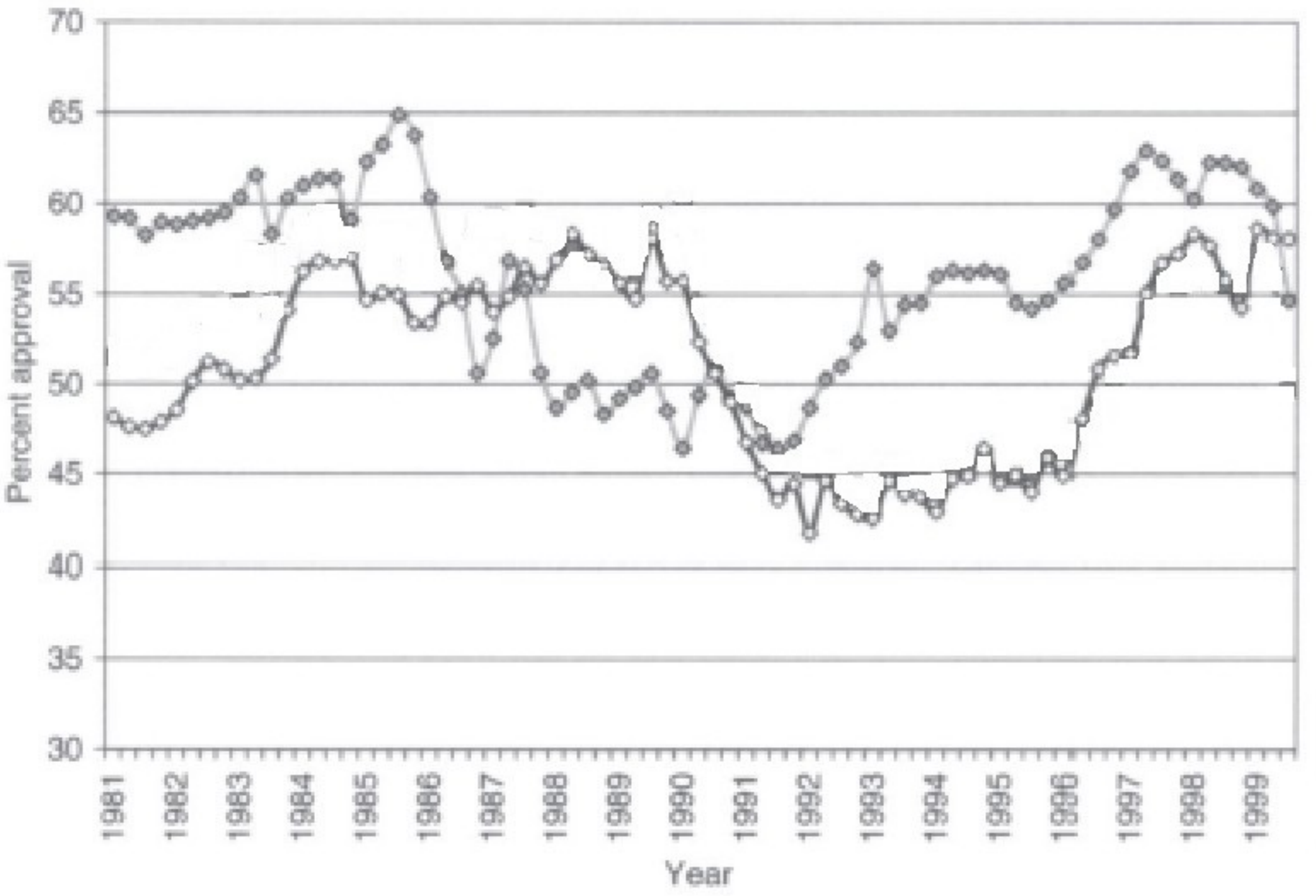

Fig. 5. Democratic (lower curve) and Republican (upper curve) Senator Approval, 1981-2000.

These assessments of political parties move macropartisanship along with respective actions by the parties' members. They implicitly include evaluations of the parties' leaders, a president or presidential candidate, which can be studied independently as Presidential Approval, their “deep states”, and the country’s economic states and/or prospects for the future (Consumer Sentiments or Expectations). In addition, they may also include an assessment of the country’s potential political “greatness” and even prospects for its simple survival.

The ongoing currently R.F.-Ukraine war, for example, may well provoke a U.S.-R.F. nuclear war, which, due to some forgotten results of numerical modelling, might be the very last war in the entire human history because of the following it so-called “Nuclear Winter”, which may result in the full destruction and disappearance of the entire human civilization. A President Trump’s approach to this risk and his promise to end this war was a very important part of his 2024 election campaign.

Altogether, these cumulative assessments may be called Partisanship Approval or Macropartisanship Sentiments. Their frequencies are so low that attraction oscillations seem to be the best choice to interpret them because free elastic oscillations are due to the power of doubts, which should not work at long character scales.

Out of two introduced here types of attraction oscillations, related to equations (10) and (11), the equation (10) is more appropriate owing to its lower frequency $\omega = \sqrt{g/R}$, because $l < R$ for considered ordinary public social bodies. As it is noticed in the second section, this frequency is the same for all of these bodies because both $g$ and $R$ do not depend on a particular body, and this kind of attraction oscillations is called “universal” ones.

Therefore, the “free-fall” acceleration $g$ can be approximately evaluated from the relation

$$g = R \cdot 4\pi^2 / T^2 . \qquad (13)$$

A very rough, tentative estimate of $R$ for the U.S. is just a number of its registered voters, i.e., about 186.5 million. For the sake of generality, it can be expressed in *leo* units intended to measure dimensions in the Bayesian space. Then $R = 186.5 \cdot 10^6 leo$, T = 13 years = 4,745 days, and it is easy to obtain from the (13) that $g = 327$ *leo/day*$^2$ for the United States. The oscillations described by the equation (10) refer to the case of a social body oscillating along a straight line in Fig. 1 and professional assessments do not depend on ethical ones.

The equation (11) is intended for a case where variabilities of one of two assessments (e.g. professional in Fig. 1) depend on each other explicitly or implicitly. The smaller $l$, the higher the oscillation frequency. These motions are called here “pendulum-like” oscillations. If (13) works, and the above-assessed $g = 327$ *leo/day*$^2$ is the case, then

$$l = g \cdot T^2/4\pi^2. \tag{14}$$

While considered period of 13 years is the longest one, shorter periods of about four years are also seen in Fig. 4 for Macropartisanship and in Fig. 5. If $T = 4$ years, and the oscillations are really pendulum-like attraction ones, it is easy to obtain from (14) that $l = 17.7 \cdot 10^6$*leo* and $R/l = 10.6$, i.e., $l$ is indeed smaller than $R$.

If low-frequency oscillations of Presidential Approval (Fig. 2 for Trump and Fig. 3 for Obama) are indeed attraction oscillations and T is equal to 42 days for Trump (two full periods in Fig. 2 from 05.11.17 to 28.01.18) and 32 days for Obama (one full period in Fig. 3 from 07.02.16 to 20.03.16), then $l_O$=8,490*leo*, $l_T$ =14,626*leo*, and $l_T = 1.72 l_O$. Therefore, the public assessment of Donald Trump’s professional performance was less dependent on his implicit ethical assessment than that of Obama’s.

## CONCLUSION

Deterministic social dynamics (Zhdanov M., 2024), drawn on the existence and frequent manifestations of fluctuations in various areas of social life as well as on the concepts of market power (Danov M.A., Smith J.B., and Mitchell R.K., 2003) and transaction triangle (Mitchell R.K., 2001, 2002), is applied to analyze U.S. federal, including presidential, elections, and the nature of Macropartisanship (MacKuen M.B., Erikson A.S., and Stimson J.A., 1989), i.e., what moves macropartisanship. Fluctuations of survey data on Presidential Approval, Consumer Sentiments/Expectations, and U.S. Senator Approval are also interpreted in terms of social dynamics.

It is shown that weights used to differentiate states' voting power in U.S. federal elections are a limiting case, under some very strong assumptions, of the states' masses, a concept of presented here social dynamics.

It is also shown that Bayesian space of assessments, an important part of social dynamics, is a generalization of concepts of rates, ratings, etc., widely used to measure public approval of various social bodies, including individuals, and that the space of ratings, for example, is a truncated version of the Bayesian space of assessments. Therefore, equations of social dynamics are applicable to results of popular methods of social observations.

Solutions of these equations are simple social harmonic oscillations. They are very important, despite their simplicity, because virtually any periodic function can be decomposed into a Fourie series, i.e., represented as a sum of simple harmonic oscillations. Frequencies of these oscillations are their own, natural frequencies, independent on forces, acting on social bodies and activating the oscillations. That is why, "Life is Not Fair" (Hansen S.B., 1999), and state governors' job performance ratings are not correlated with their measured results.

The lowest observed frequencies of analyzed oscillations are related to the variability of macropartisanship, where the longest period is about 13 years. Its frequency is interpreted as a natural frequency of universal attraction oscillations, independent of oscillating body parameters and defined by evaluator parameters only. If the evaluator is the United States as a whole, its obtained free-fall acceleration is $g$ =327 $leo/day^2$ under an assumption that the U.S. social body's radius $R = 186.5 \cdot 10^6 leo$. I.e., it is considered equal to the number of its registered voters, which was 186.5 million in 2024.

All higher frequencies of observed fluctuations, except for the highest ones, are prescribed to pendulum-like attraction oscillations, where professional assessments are tied to other types of assessments, such as implicit ethical or cultural ones. The degree of this dependence is revealed by a value of the parameter $l$, which relative estimates are made in the very last section right before the Conclusion for a couple of considered social bodies.

Finally, the highest observed frequencies with periods of 14, 21, and 28 days are prescribed here to free social elastic oscillations. Their analysis allows estimating relative coefficients of elasticity for different evaluators of the same social body and relative masses of social bodies evaluated by the same assessor. Estimates of relative masses are made in the subsection Presidential Approval for two social bodies evaluated in reference to the United States.

REFERENCES

Anderson J.L. and Newmark A.J., 2002. A Dynamic Model of U.S. Senator Approval, 1981-2000. *State Politics and Policy Quarterly*, 2(3):298-316.

Asrian V., Fuchs W., and Green B., 2019. Liquidity Sentiments. *The American Economic Review*, Vol. 109, No. 11, pp. 3813-3848.

Bidah S., Zakary O., Rachik M., and Ferjouchia H., 2020. Mathematical Modelling of Public Opinion: Parameter Estimation, Sensitivity Analysis, and Model Uncertainty Using an Agree-Disagree Opinion Model. *Hindawi Abstract and Applied Analysis*, 15 pages, 10.1155/2020/1837364.

Danov M., 2024. On the Nature of Social Oscillations. *Research Gate:* researchgate.net/publication/387744796.

Danov, M. A., Smith, J.B., Mitchell, R.K. 2003. Relationship prioritization for technology commercialization. *Journal of Marketing Theory and Practice*, 11(3): 59-70.

Green D., Palmquist B., and Schickler E., 1998. Macropartisanship: A Replication and Critique. *The American Political Science Review*, 92(4):883-899.

Hansen S.B., 1999. “Life is Not Fair”: Governors’ Job Performance Ratings and State Economies. *Political Research Quarterly*. 52(1):167-188.

Kittel C., Knight W.D., and Ruderman M.A., 1973 (2nd Ed.). *Berley Physics Course. Mechanics.* Vol.1, Chapter 9, p. 270.

MacKuen M.B., Erikson R.S., and Stimson J.A., 1989. Macropartisanship. *American Political Science Review*, 83 (December): 1125-42.

Martin, F. 2022. Organizational virtues and organizational anthropomorphism. *Journal of Business Ethics*, 177(1): 1-17.

Mitchell, R. K. 2001. *Transaction cognition theory and high performance economic results* (1st ed.). Victoria, BC: International Center for Venture Expertise.

Mitchell, R. K. 2002. Entrepreneurship and stakeholder theory: Comment on Ruffin Lecture #2 delivered by Professor S. Venkataraman. In: *Ruffin Series in Business Ethics. Ethics and Entrepreneurship*, vol. 3: 175-195. Charlottesville, VA: Society for Business Ethics.

Sepinwall, A. 2019. Corporations are people too (And they should act like it), by Kent Greenfield. New Haven, CT: Yale University Press, 2018. 296 pp. - We the corporations: how American businesses won their civil rights, by Adam Winkler. New York: W.W.Norton, 2018. 496 pp. *Business Ethics Quarterly*, 29(4): 550-554.

Tolstoy, L. 2010. *War and Peace*. New York: Oxford University Press.

Zhdanov M., 2024. Idealized Social Dynamics in Bayesian Space of Assessments. *Arxiv*, *arXiv:2406.08432*